\documentclass[12pt]{article}

\usepackage{a4wide}
\usepackage{amsmath}
\usepackage{amsthm}
\usepackage{amsfonts,amssymb}

\makeatletter
\@addtoreset{equation}{section}

\makeatother

\def\e{\mbox{\rm e}}

\def\Define{:=}

\def\nn{\nonumber}

\newcommand{\myref}[1]{(\ref{#1})}
\newtheorem{thm}{Theorem}[section]
\newtheorem{prop}[thm]{Proposition}
\newtheorem{lm}[thm]{Lemma}
\newtheorem{cor}[thm]{Corollary}
\newtheorem{defn}[thm]{Definition}

\title{\Large\bf
Bosonic and Fermionic Eigenstates \\
for Generalized Sutherland Models
}

\author{
Akinori Nishino\footnote{E-mail address: %
nishino@monet.phys.s.u-tokyo.ac.jp}
and Miki Wadati \\
\,\\
Department of Physics, Graduate School of Science,\\
University of Tokyo,\\
Hongo 7--3--1, Bunkyo-ku, Tokyo 113--0033, Japan\\
}

\begin{document}

\setlength{\baselineskip}{19pt}

\maketitle
\begin{center}
\underline{{ABSTRACT}}
\end{center}
We construct bosonic and fermionic eigenstates 
for the generalized Sutherland models
associated with arbitrary reduced root systems respectively,
through $W$-symmetrization and $W$-anti-symmetrization
of Heckman-Opdam's nonsymmetric Jacobi polynomials. 
Square norms of the nonsymmetric Heckman-Opdam polynomials
are evaluated from their Rodrigues formulae.
The $W$-symmetrization and $W$-anti-symmetrization 
of the nonsymmetric polynomials
enable us to evaluate square norms of 
bosonic and fermionic eigenstates for the generalized Sutherland models.

\section{Introduction}
In 1971, Sutherland introduced a quantum many-body system
on a unit circle ($0\leq \theta_{j}<2\pi$)
\cite{Sutherland_71PRA,Sutherland_72PRA},
\begin{align}
 \label{eq:Sutherland1971}
 H\Define 
  -\frac{1}{2}\sum_{j=1}^{N}
  \frac{\partial^{2}}{\partial \theta_{j}^{2}}
  +\!\!\sum_{1\leq j< k\leq N}\!
   \frac{g}{\sin^{2}(\theta_{j}-\theta_{k})},
\end{align}
which is now called the Sutherland model.
The model has the same number of independent and 
mutually commutative conserved operators as its degrees of freedom,
and therefore is a quantum integrable system.
The conserved operators have joint eigenvectors which
can be expressed by products of the Jastrow type wave function and 
the Jack polynomials.
The Jack polynomials not only enable us to calculate
the exact correlation functions of the model \cite{Ha_94PRL,Uglov_98CMP}
but also provide powerful tools in the theory of
condensed matter physics \cite{Kato_98PRL}.

Quantum mechanical systems
which describe many particles with inverse-square-type long-range
interactions in a one-dimensional space are, in general, called 
the Calogero-Sutherland (CS) models~\cite{Calogero_71JMP,Sutherland_71PRA,
Sutherland_72PRA,Olshanetsky-Perelomov_83PR}.
A systematic construction of the
commutative conserved operators of the CS models
are provided from
the Dunkl-Cherednik operator formulations 
\cite{Cherednik_91InvMath,Dunkl_89TAMS,Opdam_98mathRT}.
Generalizations of the formulations
are applied to a wide class of the CS models
and clarify the relationships with other integrable models
\cite{Hikami-Wadati_97Topics,Komori-Hikami_98LMP,
Komori_98LMP,Komori_99preprint}.
Among them, the CS models with trigonometric interactions,
for example, the Sutherland model \myref{eq:Sutherland1971}, are
studied in the context of the double affine Hecke algebras.
The double affine Hecke algebras were introduced by
Cherednik to reconstruct the theories of the Macdonald polynomials
\cite{Cherednik_95AnnMath,
Cherednik_97SelectaMath,Macdonald,Macdonald_95Bourbaki}.
In the differential setting \cite{Opdam_98mathRT}, they also
give a unified treatment of the conserved operators
and the orthogonal polynomials appearing in their eigenstates.
Lapointe and Vinet introduced the raising operators which
create the bosonic eigenstates of the Sutherland model~\cite{
Lapointe-Vinet_95IMRN,Lapointe-Vinet_97AdvMath}.
As for mathematics, they presented the Rodrigues formula
for the Jack polynomial
and proved integrality of its coefficients.
But its extension to the Sutherland models associated with
other root systems has not been established.

In the previous papers \cite{Nishino-Ujino_99NPB,
Nishino-Komori_00preprint},
we presented an algebraic construction of the nonsymmetric Macdonald
polynomials and evaluated square norms of the Macdonald polynomials
through symmetrization of scalar products of the 
nonsymmetric Macdonald polynomials.
In this paper,
we consider bosonic and fermionic eigenstates 
for the generalized Sutherland models associated with arbitrary reduced
root systems through $W$-symmetrization and $W$-anti-symmetrization of
the nonsymmetric Heckman-Opdam polynomials.
This paper is organized as follows: in Section 2, we briefly describe
the affine root systems and the
degenerate double affine Hecke algebras following Opdam and Cherednik.
The commutative Dunkl-Cherednik operators are introduced.
In Section 3, we provide the Rodrigues formulae for the 
nonsymmetric Heckman-Opdam polynomials and evaluate their square norms.
By use of $W$-symmetrization and $W$-anti-symmetrization method,
we algebraically construct the bosonic and fermionic eigenstates
for the generalized Sutherland models and evaluate their square norms
in Section 4.
The final section is devoted to concluding remarks.

\section{Degenerate Double Affine Hecke Algebras}
\subsection{Extended Affine Weyl Groups}
We start with the definition of the extended affine Weyl group
which acts on the affine coroot system \cite{Humphreys,Cherednik_95AnnMath}.
Let $V$ be an $N$-dimensional real vector space with a positive
definite symmetric bilinear form $\langle\cdot,\cdot\rangle$.
Let $R\subset V$ be an irreducible reduced root system
which corresponds to the simple Lie algebra 
of type $A, B, C, D, E, F$ and $G$.
We take a root basis $\Pi=\{\alpha_{i}|i\in I\}$ of $R$
where $I=\{1,2,\cdots,N\}$ a set of indices.
A decomposition of $R$ is fixed by the following disjoint union:
$R=R_{+}\cup R_{-}$, where $R_{+}$ is the set of positive roots
relative to $\Pi$ and $R_{-}=-R_{+}$.
We denote by $R^{\vee}(\subset V)$ the coroot system which has the elements
$\alpha^{\vee}\Define 2\alpha/\langle\alpha,\alpha\rangle$
corresponding to the roots $\alpha\in R$.
Let $\Pi^{\vee}=\{\alpha^{\vee}|\alpha\in \Pi\}$ be a root basis 
of $R^{\vee}$.
We define the fundamental weights $\Lambda_{i}$
and coweights $\Lambda_{i}^{\vee}$
such that $\langle \alpha_{i}^{\vee},\Lambda_{j}\rangle=\delta_{ij}$
and $\langle \Lambda_{i}^{\vee},\alpha_{j}\rangle=\delta_{ij}$
respectively.
We use the standard notations $Q$, $Q^{\vee}$, $P$ and $P^{\vee}$
for the root lattice, the coroot lattice, the weight lattice
and the coweight lattice respectively,
\begin{align}
 &Q\Define \bigoplus_{i\in I}\mathbb{Z}\alpha_{i}
  \subset P\Define 
  \bigoplus_{i\in I}\mathbb{Z}\Lambda_{i}, \quad
  Q^{\vee}\Define \bigoplus_{i\in I}\mathbb{Z}\alpha_{i}^{\vee}
  \subset P^{\vee}\Define 
  \bigoplus_{i\in I}\mathbb{Z}\Lambda_{i}^{\vee},
\end{align}
and $Q_{+}, P_{+}, Q_{+}^{\vee}$ and $ P_{+}^{\vee}$
for the corresponding lattices with $\mathbb{Z}_{+}$
instead of $\mathbb{Z}$.
The reflection on $V$
with respect to the hyperplane orthogonal to 
$\alpha^{\vee}\in R^{\vee}$
is defined by
\begin{align}
 s_{\alpha^{\vee}}(\mu)\Define
 \mu-\langle\alpha^{\vee},\mu\rangle\alpha,\quad
 \text{ for }\, \mu\in V.
\end{align}
The reflections associated with the simple roots 
$\{s_{\alpha^{\vee}}|\alpha^{\vee}\in\Pi^{\vee}\}$, i.e.,
the simple reflections,
generate the Weyl group $W$.
The simple reflections are related to each other
by $(s_{\alpha_{i}^{\vee}}s_{\alpha_{j}^{\vee}})^{m_{ij}}=1$
where $m_{ij}=2,3,4,6$ if $\alpha_{i}^{\vee}$ and $\alpha_{j}^{\vee}$
are connected by $0,1,2,3$ laces
in the dual Dynkin diagram $\Gamma$, respectively.
The length $l$ of $w\in W$ is defined from a reduced (shortest) expression
$w=s_{j_{l}}\cdots s_{j_{2}}s_{j_{1}}$.
We denote the set of distinct weights lying 
in the $W$-orbit of $\mu\in P$ by $W(\mu)$
and a unique dominant weight in $W(\mu)$ by $\mu^{+}(\in P_{+})$.
We define the order $\leq$ on $P$ by
\[
 \nu\leq\mu, \,(\mu, \nu\in P)\quad \Leftrightarrow \quad
 \mu-\nu\in Q_{+}.
\]

We turn to the affine coroot system
$\widehat{R}^{\vee}\Define R^{\vee}\times \mathbb{Z}K \subset 
 \widehat{V}\Define V\oplus \mathbb{R}K$.
Let $\widehat{R}_{+}
     =\{\alpha^{\vee}+k K|\alpha\in R^{\vee},k>0\}
      \cup\{\alpha^{\vee}\in R_{+}^{\vee}\}$ be
the set of positive affine coroots and
let $\hat{I}=\{0,1,\cdots,N\}$ be a set of indices
Let $\theta^{\vee}$ be the highest root in $R^{\vee}$ and
$\widehat{\Pi}^{\vee}=\{\alpha_{i}^{\vee}|i\in \hat{I}\}$
the root basis of $\widehat{R}^{\vee}$
where $\alpha_{0}^{\vee}\Define -\theta^{\vee}+K$.
We take a pairing on $\widehat{V}\times V$ as 
$\langle\hat{\lambda},\mu\rangle=\langle\lambda,\mu\rangle-h$
for $\lambda\Define \lambda+h K\in \widehat{V}$ and $\mu\in V$
and define the fundamental alcove
$C=\{\mu\in V|
\langle\alpha^{\vee},\mu\rangle<0, 
\alpha^{\vee}\in \widehat{\Pi}^{\vee}\}$.
The affine reflection on $\widehat{V}$
relative to 
$\hat{\alpha}^{\vee}=\alpha^{\vee}+k K\in \widehat{R}^{\vee}, 
(\alpha^{\vee}\in R^{\vee})$
is defined by
\[
 s_{\hat{\alpha}^{\vee}}(\hat{\lambda})\Define
 \hat{\lambda}-\langle\lambda,\alpha^{\vee}\rangle\hat{\alpha}^{\vee},
 \quad \text{ for }\,
 \hat{\lambda}=\lambda+h K\in\widehat{V},
\]
which induces the dual action on $V$ 
through the pairing $\langle\cdot,\cdot\rangle$,
\[
 s_{\hat{\alpha}^{\vee}}\langle\mu\rangle=
 \mu-(\langle\alpha^{\vee},\mu\rangle-k)\alpha,
 \quad \text{ for }\, \mu\in V.
\]
The affine reflections associated with the simple roots 
$\{s_{\alpha}^{\vee}|\alpha^{\vee}\in\widehat{\Pi}^{\vee}\}$ 
generate the affine Weyl group $\widehat{W}$.
Let $\tau_{\nu}, (\nu\in P)$ be the translation
on $\widehat{V}$,
\[
 \tau_{\nu}(\hat{\lambda})
 \Define\hat{\lambda}-\langle\lambda,\nu\rangle\delta,
 \quad \text{ for }\, \hat{\lambda}=\lambda+h K\in \widehat{V},
\]
which induces the translation on $V$,
\[
 \tau_{\nu}\langle\mu\rangle
 =\mu-\nu,
 \quad \text{ for }\, \mu\in V.
\]
One sees that
the elements $\{\tau_{\kappa}|\kappa\in P^{\vee}\}$ 
are mutually commutative. 
The affine Weyl group $\widehat{W}$ contains the element 
$\tau_{\alpha}= s_{-\alpha^{\vee}+K}s_{\alpha^{\vee}}, (\alpha\in Q)$
which is interpreted as a translation 
corresponding to the root $\alpha\in Q$.
In fact, the affine Weyl group is isomorphic to the semidirect product
$\widehat{W}\simeq W\ltimes\tau_{Q}$.

We define the extended affine Weyl group 
by a semidirect product $\widetilde{W}\Define W\ltimes\tau_{P}$.
One finds that $\widetilde{W}$
is defined so as to preserve the affine coroot systems $\widehat{R}^{\vee}$.
Let $\Omega\Define\{\tilde{w}\in \widetilde{W}|\tilde{w}(C)=C\}$.
The extended affine Weyl group is isomorphic to a
semidirect product $\widetilde{W}\simeq \Omega\ltimes \widehat{W}$.
Let ${\cal O}$ be a set of indices of the image of $\alpha_{0}^{\vee}$ 
by the automorphism of the dual Dynkin diagram $\Gamma$.
A weight $\mu\in P_{+}$ satisfying 
$0\leq\langle\alpha^{\vee},\mu\rangle\leq 1$
for every $\alpha^{\vee}\in R_{+}^{\vee}$ is called a minuscule weight.
It is known that the set of minuscule weights is given by 
$\{\Lambda_{r}|r\in{\cal O}\}$ where we put $\Lambda_{0}=0$
(see, for example, \cite{Cherednik_95AnnMath}).
One sees a decomposition $\tau_{\Lambda_{r}}=\omega_{r}w_{r}$
with $\omega_{r}\in \Omega$ and $w_{r}\in W$
if and only if $r\in {\cal O}$.
Each $\omega_{r}$
is distinguished by $\omega_{r}(\alpha_{0})=\alpha_{r}$.
Note that $\omega_{0}=w_{0}=1$.
Let $r^{*}\in {\cal O}$ be the index such that 
$\alpha_{r^{*}}=\omega_{r}^{-1}(\alpha_{0})$.

We extend the definition of the length
to an element $\tilde{w}\in\widetilde{W}$ as
\begin{align}
 \label{eq:length_w}
 l(\tilde{w})\Define|R_{\tilde{w}}^{\vee}|, \, \text{ where } \,
 R_{\tilde{w}}^{\vee}\Define
 \widehat{R}_{+}^{\vee}\cap\tilde{w}^{-1}\widehat{R}_{-}^{\vee},
\end{align}
which is consistent with that for $w\in W$,
that is $l=l(w)$.
Here $R_{\tilde{w}}^{\vee}$ is the set of positive coroots
which become negative coroots by the action of $\tilde{w}$.
If we take a reduced expression of $\tilde{w}\in\widetilde{W}$ as
$\tilde{w}=\omega_{r}s_{i_{l}}\cdots s_{i_{2}}s_{i_{1}}$,
the set $R_{\tilde{w}}^{\vee}$ is explicitly given by
\begin{align}
 \label{eq:R_w}
 R_{\tilde{w}}^{\vee}
 =\big\{\alpha^{(1)}=\alpha_{i_{1}},
        \alpha^{(2)}=s_{i_{1}}(\alpha_{i_{2}}),\cdots,
        \alpha^{(l)}=s_{i_{1}}s_{i_{2}}\cdots s_{i_{l\!-\!1}}
                     (\alpha_{i_{l}})\big\},
\end{align}
which is independent of the decomposition of $\tilde{w}$.
For $\tilde{w}=\tau_{\lambda}$, we have
\begin{align}
 \label{eq:R_tau}
 R_{\tau_{\mu}}^{\vee}
 =\{\alpha^{\vee}+k K|\alpha^{\vee}\in R_{+}^{\vee},
    \langle\alpha^{\vee},\mu\rangle>k\geq 0\}
  \cup
  \{\alpha^{\vee}+k K|\alpha^{\vee}\in R_{-}^{\vee},
    \langle\alpha^{\vee},\mu\rangle\geq k>0\}.
\end{align}

We take a set of parameters 
$\{k_{\alpha}\in \mathbb{C}|\alpha\in R\}$
such that $k_{\alpha}=k_{w(\alpha)}$ for $w\in W$.
We write $k_{i}=k_{\alpha_{i}}, (i\in I)$.
Define 
\begin{align}
  \rho=\frac{1}{2}\sum_{\alpha\in R_{+}}\alpha,\quad
  \rho_{k}=\frac{1}{2}\sum_{\alpha\in R_{+}}k_{\alpha}\alpha. 
\end{align}
Let $P_{++}\Define P_{+}+\rho$ be the regular dominant weight lattice.
We see $\langle\alpha^{\vee},\mu\rangle\gneq 0$ for $\mu\in P_{++}$ 
and $\alpha^{\vee}\in \Pi^{\vee}$.

\subsection{Degenerate Double Affine Hecke Algebras}

Following Opdam and Cherednik~\cite{Cherednik_91InvMath,
Cherednik_97SelectaMath,Opdam_98mathRT},
we introduce degenerate double affine Hecke algebras ${\cal DH}$.
The commutative elements of ${\cal DH}$ give the
commutative differential operators in 
${\rm End}(\mathbb{C}[P])$
which are referred to as the Dunkl-Cherednik 
operators~\cite{Cherednik_91InvMath,Dunkl_89TAMS}
and uniquely characterize Heckman-Opdam's nonsymmetric Jacobi polynomials
as their eigenvectors.

\begin{defn}[degenerate double affine Hecke algebra]
 The degenerate double affine Hecke algebra ${\cal DH}$
 is generated over the field $\mathbb{C}$ by the elements
 $\{s_{i},\omega_{r},D^{\Lambda_{j}^{\vee}}|
    i\in \hat{I},r\in {\cal O},j\in I\}$ satisfying
\begin{align}
 \text{i)}&\quad
  (s_{i}s_{j})^{m_{ij}}=1,\,
  \text{ for }\, 0\leq i,j\leq N, \nn\\
 \text{ii)}&\quad
  \omega_{r}s_{i}\omega_{r}^{-1}=s_{j},\,
  \text{ if }\, \omega_{r}(\alpha_{i}^{\vee})=\alpha_{j}^{\vee},\,
  \text{ for }\, r\in {\cal O}, \nn\\
 \text{iii)}&\quad
  D^{\lambda}D^{\mu}=D^{\mu}D^{\lambda},\,
  \text{ for }\, \lambda,\mu\in P^{\vee},\nn\\
 \text{iv)}&\quad
  s_{i}D^{\lambda}=D^{\lambda}s_{i}, 
  \text{ if }\,  \langle\lambda,\alpha_{i}\rangle=0,\,
  \text{ for }\, 1\leq i\leq N, \nn\\
  &\quad
  s_{0}D^{\lambda}=D^{\lambda}s_{0}, 
  \text{ if }\,  \langle\lambda,-\theta\rangle=0, \nn\\
 \text{v)}&\quad
  s_{i}D^{\lambda}-D^{s_{i}(\lambda)}s_{i}=k_{i}, 
  \text{ if }\,  \langle\lambda,\alpha_{i}\rangle=1,\,
  \text{ for }\, 1\leq i\leq N, \nn\\
  &\quad
  s_{0}D^{\lambda}-D^{s_{0}(\lambda)}s_{0}=k_{\theta}, 
  \text{ if }\,  \langle\lambda,-\theta\rangle=1, \nn\\
 \text{vi)}&\quad
  \omega_{r}D^{\lambda}\omega_{r}^{-1}
  =D^{\omega_{r}(\lambda)}
  =D^{w_{r}^{-1}(\lambda)}-\langle\lambda,\Lambda_{r^{*}}\rangle,\,
  \text{ for }\, r\in {\cal O},
\end{align}
where 
\[
 D^{\hat{\lambda}}=\sum_{i\in I}\lambda_{i}D^{\Lambda_{i}^{\vee}}-h,\,
 \text{ for }\,
 \hat{\lambda}
 =\sum_{i\in I}\lambda_{i}\Lambda_{i}^{\vee}+h K.
\]
\end{defn}
The degenerate double affine Hecke algebra ${\cal DH}$ contains
the extended affine Weyl group $\widetilde{W}$
which acts on the affine coroot system $\widehat{R}^{\vee}$.
Repeated use of the defining relations of ${\cal DH}$
gives the relations,
\begin{align}
 &s_{i}D^{\lambda}-D^{s_{i}(\lambda)}s_{i}
 =k_{i}\langle\lambda,\alpha_{i}\rangle,\,
 \text{ for }\, 1\leq i\leq N, \nn\\
 &s_{0}D^{\lambda}-D^{s_{0}(\lambda)}s_{0}
 =k_{\theta}\langle\lambda,-\theta\rangle.
\end{align}
We define the following elements:
\[
  X^{-\mu}\Define\tau_{\mu}\in {\cal DH},\,
  \text{ for }\, \mu\in P,
\]
which provide $s_{0}=X^{-\theta}s_{\theta^{\vee}}$
and $\omega_{r}=X^{-\Lambda_{r}}w_{r}^{-1}, (r\in{\cal O})$.

We introduce the commutative differential operators
$\{\hat{D}^{\hat{\lambda}}\in {\rm End}(\mathbb{C}[P])|
 \hat{\lambda}=\lambda+h K\in \widehat{P}^{\vee}\}$,
\begin{align}
 \label{eq:DC-op}
 \hat{D}^{\hat{\lambda}}f\Define
 \partial^{\lambda}(f)
 +\sum_{\alpha\in R_{+}}
  \frac{k_{\alpha}\langle\lambda,\alpha\rangle}
       {x^{\alpha}-1}(f-\hat{s}_{\alpha}(f))
 +\langle\lambda,\rho_{k}\rangle f-h f,
\end{align}
where $\{\partial^{\lambda}\in {\rm End}(\mathbb{C}[P])
|\lambda\in P^{\vee}\}$
is the derivative of $\mathbb{C}[P]$:
\[
 \partial^{\lambda}(x^{\mu})=\langle\lambda,\mu\rangle x^{\mu},
 \quad \text{ for }\, x^{\mu}\in \mathbb{C}[P],
\]
and the elements $w\in W$
act on $\mathbb{C}[P]$ as
\[
  w(x^{\mu})=x^{w(\mu)},\quad
  \text{ for } x^{\mu}\in\mathbb{C}[P].
\]
We consider $x^{\mu}, (\mu\in P)$
as operators of multiplication of $x^{\mu}$.
The differential operators $\{\hat{D}^{\lambda}|\lambda\in P^{\vee}\}$
are called the Dunkl-Cherednik 
operators~\cite{Cherednik_91InvMath,Dunkl_89TAMS}.
One sees that
a map $\pi :{\cal DH}\rightarrow 
{\rm End}(\mathbb{C}[P])$
defined by
\begin{align}
 \pi: s_{i}\mapsto s_{i},\, (1\leq i\leq N),\quad
      D^{\hat{\lambda}}\mapsto \hat{D}^{\hat{\lambda}},\quad
      X^{\mu}\mapsto x^{\mu},
\end{align}
gives a faithful representation of ${\cal DH}$.
The Dunkl-Cherednik operators 
$\{\hat{D}^{\lambda}|\lambda\in P^{\vee}\}$
have the triangularity in $\mathbb{C}[P]$:
\begin{align}
 \label{eq:D_triangle}
 \hat{D}^{\lambda}x^{\mu}
 =\langle\lambda,\mu+\rho_{k}(\mu)\rangle x^{\mu}
  +\sum_{\nu\prec\mu}c_{\mu\nu}x^{\nu},
 \quad \mu\in P,\,\, c_{\mu\nu}\in \mathbb{K},
\end{align}
where we denote by $w_{\mu}$ the shortest element
of $W$ such that $w_{\mu}^{-1}(\mu)\in P_{+}$ 
and define $\rho_{k}(\mu)\Define w_{\mu}(\rho_{k})$.
The order $\preceq$ on $P$ is defined by
\begin{align}
 \nu\preceq \mu,\, (\mu,\nu\in P)
 \Leftrightarrow\left\{
 \begin{array}{l}
  \text{ if }\, \mu^{+}\neq\nu^{+},\,
  \text{ then }\, \nu^{+}<\mu^{+}, \\
  \text{ if }\, \mu^{+}=\nu^{+},\,
  \text{ then }\, \nu\leq\mu.
 \end{array}
 \right.
\end{align}

\section{Nonsymmetric Heckman-Opdam polynomials}

We investigate the eigenvectors of the Dunkl-Cherednik 
operators $\{\hat{D}^{\lambda}|\lambda\in P^{\vee}\}$.
Due to the triangularity \myref{eq:D_triangle},
there exists a family of polynomials 
$F_{\mu}\Define F_{\mu}(x;\{k_{\alpha}\})\in \mathbb{C}[P], (\mu\in P)$
satisfying the following conditions:
\begin{align}
 \label{eq:nonsym-HO_def}
 \text{i)}&\quad
  F_{\mu}=x^{\mu}+\sum_{\nu\prec\mu}w_{\mu\nu}x^{\nu}, \quad
  w_{\mu\nu}\in \mathbb{C}, \nn\\
 \text{ii)}&\quad
  \hat{D}^{\lambda}F_{\mu}
  =\langle\lambda,\mu+\rho_{k}(\mu)\rangle F_{\mu}.
\end{align}
$F_{\mu}$ are the nonsymmetric Jacobi polynomials 
introduced by Heckman and Opdam.
Hereafter
we call them the nonsymmetric Heckman-Opdam polynomials.
Note that all the eigenspaces of the Dunkl-Cherednik 
operators $\{\hat{D}^{\lambda}|\lambda\in P^{\vee}\}$ are one-dimensional.
Applying $\{s_{i}\in {\rm End}(\mathbb{C}[P])|i\in I\}$
to the nonsymmetric Heckman-Opdam polynomials $F_{\mu}, (\mu\in P)$, 
we see that
\begin{equation}
 s_{i}F_{\mu}=
 \begin{cases}
  \dfrac{k_{i}}{\langle\alpha_{i}^{\vee},\mu+\rho_{k}(\mu)\rangle}F_{\mu}
  +F_{s_{i}(\mu)}, 
  &\text{ if }\,\langle\alpha_{i}^{\vee},\mu\rangle<0,
  \\[4mm]
  F_{\mu}, 
  &\text{ if }\,\langle\alpha_{i}^{\vee},\mu\rangle=0, 
  \\[2mm]
  \dfrac{k_{i}}
         {\langle\alpha_{i}^{\vee},\mu+\rho_{k}(\mu)\rangle}F_{\mu}
  +\dfrac{\langle\alpha_{i}^{\vee},\mu+\rho_{k}(\mu)\rangle^{2}-k_{i}^{2}}
          {\langle\alpha_{i}^{\vee},\mu+\rho_{k}(\mu)\rangle^{2}}
          F_{s_{i}(\mu)},
  &\text{ if }\,\langle\alpha_{i}^{\vee},\mu\rangle>0.
 \end{cases}
\end{equation}

We introduce intertwiners in ${\cal DH}$
to provide the Rodrigues formulae for the nonsymmetric 
Heckman-Opdam polynomials.
These intertwiners were first considered in 
the context of the Hecke algebras.
The intertwiners in ${\cal DH}$ were developed by 
Cherednik~\cite{Cherednik_97SelectaMath}.
With their use,
Opdam derived the evaluation formulae of 
the nonsymmetric Heckman-Opdam polynomials~\cite{Opdam_98mathRT}.
We construct the Rodrigues formulae for the nonsymmetric 
Heckman-Opdam polynomials and algebraically
evaluate their square norms.
\begin{defn}
\begin{enumerate}
\item[i)]
We define $\{K_{i}\in {\cal DH}|i\in \hat{I}\}$ by
\begin{align}
  K_{0}\Define s_{\theta}X^{\theta}D^{\alpha_{0}^{\vee}}-k_{\theta},
  \quad
  K_{i}\Define s_{i}D^{\alpha_{i}^{\vee}}-k_{i},
\end{align}
which we call the intertwiners.
\item[ii)]
For a reduced expression 
$w=\omega_{r}s_{i_{l}}\cdots s_{i_{2}}s_{i_{1}}\in \widetilde{W}$, 
we define 
$K_{w}\Define \omega_{r}K_{i_{l}}\cdots K_{i_{2}}K_{i_{1}}(\in {\cal DH})$.
In particular, we write $B_{\mu}\Define K_{\tau_{-\mu}}$
for $\mu\in P_{+}$.
We call $\{\hat{B}_{\mu}\Define \pi(B_{\mu})\in {\rm End}(\mathbb{C}[P])\}$
the raising operators.
\end{enumerate}
\end{defn}
The elements $\{K_{w}|w\in \widetilde{W}\}$
have the following relations:
\begin{align}
 \label{eq:prop-int}
 \text{i)}&\quad
  K_{i}K_{j}K_{i}\cdots=K_{j}K_{i}K_{j}\cdots,\,
  m_{ij} \text{ factors on each side, } \nn\\
 &\quad
  \omega_{r}K_{i}\omega_{r}^{-1}=K_{j},\,
  \text{ if }\, \omega_{r}(\alpha_{i}^{\vee})=\alpha_{j}^{\vee}, \nn\\
 \text{ii)}&\quad
  K_{i}^{2}=-(D^{\alpha_{i}^{\vee}})^{2}+k_{i}^{2}, \,
  \text{ for }\, i\in \hat{I}, \nn\\
 \text{iii)}&\quad
  K_{w}D^{\lambda}=D^{w(\lambda)}K_{w}, \,
  \text{ for }\, \lambda\in P^{\vee}.
\end{align} 
The first relations in \myref{eq:prop-int} are called the braid relations.
The last relations give the reason why they are called intertwiners.
For the elements $\{B_{\mu}\in {\cal DH}\}$, we can show that
\begin{align}
 B_{\mu}D^{\lambda}=D^{\tau_{-\mu}(\lambda)}B_{\mu}
 =(D^{\mu}-\langle\lambda,\mu\rangle)B_{\mu}.
\end{align}
By applying the raising operators $\hat{B}_{\mu}$
to the nonsymmetric Heckman-Opdam polynomials $F_{\nu}, (\nu\in P_{+})$, 
we obtain
\begin{align}
 \hat{D}^{\lambda}(\hat{B}_{\mu}F_{\nu})
 =\hat{B}_{\mu}(\hat{D}^{\lambda}+\langle\lambda,\mu\rangle)F_{\nu}
 =\langle\lambda,\mu+\nu+\rho_{k}\rangle(\hat{B}_{\mu}F_{\nu}).
\end{align}
Hence $\hat{B}_{\mu}F_{\nu}$ coincides with $F_{\mu+\nu}$
up to a constant factor.
If we apply $\hat{K}_{w}\Define\pi(K_{w}), (w\in W)$ 
to $F_{\mu}, (\mu\in P_{+})$,
we see that
\begin{align}
 \hat{D}^{\lambda}(\hat{K}_{w}F_{\mu})
 =\hat{K}_{w}\hat{D}^{w^{-1}(\lambda)}F_{\mu}
 =\langle w^{-1}(\lambda),\mu+\rho_{k}\rangle \hat{K}_{w}F_{\mu}
 =\langle \lambda,w(\mu+\rho_{k})\rangle(\hat{K}_{w}F_{\mu}).
\end{align}
Hence $\hat{K}_{w}F_{\mu}$ coincides with $F_{w(\mu)}$
up to a constant factor.

\begin{thm}[Rodrigues formulae]
\begin{enumerate}
\item[i)]
For a dominant weight $\mu\in P_{+}$,
we construct the nonsymmetric Heckman-Opdam
polynomials $F_{\mu}$ by applying the raising operators
$\{\hat{B}_{\mu}|\mu\in P_{+}\}$ to the reference state $F_{0}=1$,
\begin{align}
 \label{eq:nonsym-HO-Rodrigues1}
 F_{\mu}=c_{\mu}^{-1}\hat{B}_{\mu} F_{0},
\end{align}
where the coefficient of the top term is given by
\[
 c_{\mu}=\prod_{\alpha^{\vee}\in R_{\tau_{-\mu}}^{\vee}}\!\!
  \langle\alpha^{\vee},\rho_{k}\rangle. 
\]
\item[ii)]
For a general weight $\mu\in P$,
we construct the nonsymmetric Heckman-Opdam polynomials $F_{\mu}$
by applying the operator $\hat{K}_{w_{\mu}}$ 
($w_{\mu}\in W,w_{\mu}^{-1}(\mu)\Define\mu^{+}\in P_{+}$)
to $F_{\mu^{+}}$,
\begin{align}
 \label{eq:nonsym-HO-Rodrigues2}
 F_{\mu}=c_{w_{\mu}}^{-1}\hat{K}_{w_{\mu}}F_{\mu^{+}},
\end{align}
where the coefficient of the top term is
\[
  c_{w_{\mu}}
  =\prod_{\alpha^{\vee}\in R_{w_{\mu}}^{\vee}}
   \frac{\langle\alpha^{\vee},\mu^{+}+\rho_{k}\rangle^{2}-k_{\alpha}^{2}}
   {\langle\alpha^{\vee},\mu^{+}+\rho_{k}\rangle}.
\]
\end{enumerate}
\end{thm}
See our previous paper \cite{Nishino-Komori_00preprint}
for the detailed proofs for the coefficients of the top terms
appearing in the Rodrigues formulae.

In the remainder of this paper,
we assume 
$k_{\alpha}\geq 0, (\alpha\in R)$.
Define the inner product $\langle\cdot,\cdot\rangle_{k}$ by
\begin{align}
 \label{eq:HO_inner-prod}
  \langle f,g\rangle_{k}\Define
  \int_{T}f(t)\overline{g(t)}\Delta_{k}(t) d \mu,
\end{align}
where $T=V/2\pi Q^{\vee}$ is a torus,
$x^{\mu}(t)\Define \e^{\sqrt{-1}\langle t,\mu\rangle}, (t\in T)$,
$d \mu$ is the normalized Haar measure on $T$
and the weight function $\Delta_{k}$ is given by
\begin{align}
  \Delta_{k}\Define
  \prod_{\alpha\in R}|1-x^{\alpha}|^{k_{\alpha}}.
\end{align}
The square norm of the reference state is given by
\begin{align}
 \langle 1,1\rangle_{k}
 =\prod_{\alpha\in R_{+}}
  \frac{\Gamma(\langle\alpha^{\vee},\rho_{k}\rangle+k_{\alpha}+1)
        \Gamma(\langle\alpha^{\vee},\rho_{k}\rangle-k_{\alpha}+1)}
       {\Gamma(\langle\alpha^{\vee},\rho_{k}\rangle+1)^{2}},
\end{align}
which is indeed evaluated from Opdam's shift operators
\cite{Opdam_89InvMath}.
Since
the Dunkl-Cherednik operators $\{\hat{D}^{\lambda}\}$ are selfadjoint 
with respect to the inner product $\langle\cdot,\cdot\rangle_{k}$ 
\myref{eq:HO_inner-prod}, we have the orthogonality,
\begin{align}
  \label{eq:nonsym-HO-orth}
  \langle F_{\mu},F_{\nu}\rangle_{k}=0,
   \text{ if } \mu\neq\nu.
\end{align}
In fact,
the nonsymmetric Heckman-Opdam polynomials form an orthogonal basis
in $\mathbb{C}[P]$ with respect to the inner product 
$\langle\cdot,\cdot\rangle_{k}$ \myref{eq:HO_inner-prod}.
We see that the adjoint operators of 
$\{\omega_{r},\hat{K}_{i}|r\in O,i\in \hat{I}\}$
are given by
\begin{align}
 \omega_{r}^{*}=\omega_{r}^{-1},\quad
 \hat{K}_{i}^{*}=-\hat{K}_{i}.
\end{align}

\begin{thm}
For a dominant weight $\mu\in P_{+}$, we have
\begin{align}
 \langle F_{\mu},F_{\mu}\rangle_{k}
 =\prod_{\alpha\in R_{+}}
  \frac{\Gamma(\langle\alpha^{\vee},\mu+\rho_{k}\rangle+k_{\alpha}+1)
        \Gamma(\langle\alpha^{\vee},\mu+\rho_{k}\rangle-k_{\alpha}+1)}
       {\Gamma(\langle\alpha^{\vee},\mu+\rho_{k}\rangle+1)^{2}}.
\end{align}
\end{thm}
\begin{proof}
Define 
$N(\hat{\alpha}^{\vee})\in {\cal DH}, 
  (\hat{\alpha}^{\vee}=\alpha^{\vee}+k K\in \widehat{R}^{\vee})$ by
\begin{align}
 N(\hat{\alpha}^{\vee})\Define 
 (D^{\alpha^{\vee}})^{2}-k_{\alpha}^{2}.
\end{align}
Since they satisfy the following properties:
\[ 
N(\hat{\alpha}_{i}^{\vee})=-K_{i}^{2},\quad
K_{w}N(\hat{\alpha}^{\vee})=N(w(\hat{\alpha}^{\vee}))K_{w},\,
\text{ for } w\in \widetilde{W},
\]
the product $\hat{B}_{\mu}^{*}\hat{B}_{\mu}\in {\rm End}(\mathbb{C}[P])$
is written as
\begin{align}
 \hat{B}_{\mu}^{*}\hat{B}_{\mu}
 &=\omega_{r}\hat{K}_{i_{l}}^{*}\cdots \hat{K}_{i_{1}}^{*}
   \hat{K}_{i_{1}}\cdots \hat{K}_{i_{l}}\omega_{r}^{*} \nn\\
 &=\omega_{r}\hat{K}_{i_{l}}^{*}\cdots \hat{K}_{i_{2}}^{*}
   \hat{N}(\alpha_{i_{1}}^{\vee})
   \hat{K}_{i_{2}}\cdots \hat{K}_{i_{l}}\omega_{r}^{-1} \nn\\
 &=\prod_{\alpha^{\vee}\in R_{\tau_{-\mu}}^{\vee}}
   \!\! \hat{N}(\alpha^{\vee}),\nn
\end{align}
where $\hat{N}(\alpha^{\vee})\Define\pi(N(\alpha^{\vee}))$.
The square norms of $F_{\mu}$ are calculated as
\begin{align}
 \langle F_{\mu},F_{\mu}\rangle_{k}
 &=\langle c_{\mu}^{-1}\hat{B}_{\mu}F_{0},
   c_{\mu}^{-1}\hat{B}_{\mu}F_{0}\rangle_{k} \nn\\
 &=(c_{\mu})^{-2}
   \langle F_{0},\hat{B}_{\mu}^{*}
   \hat{B}_{\mu}F_{0}\rangle_{k} \nn\\
 &=(c_{\mu})^{-2}
   \prod_{\alpha^{\vee}\in R_{\tau_{-\mu}}^{\vee}}
   \langle F_{0},\hat{N}(\alpha^{\vee})F_{0}\rangle_{k} \nn\\
 &=\langle F_{0},F_{0}\rangle_{k}
   \prod_{\alpha^{\vee}\in R_{\tau_{-\mu}}^{\vee}}
   \frac{\langle\alpha^{\vee},\rho_{k}\rangle^{2}-k_{\alpha}^{2}}
    {\langle\alpha^{\vee},\rho_{k}\rangle^{2}} \nn\\
 &=\langle 1,1\rangle_{k}
   \prod_{\alpha\in R_{+}}
   \prod_{i=1}^{\langle\alpha^{\vee},\mu\rangle}
   \frac{(\langle\alpha^{\vee},\rho_{k}\rangle+k+i)
         (\langle\alpha^{\vee},\rho_{k}\rangle-k+i)}
    {(\langle\alpha^{\vee},\rho_{k}\rangle+i)^{2}}. \nn
\end{align}
\end{proof}

\begin{prop}
For a weight $\mu\in P$ lying in the $W$-orbit of 
$\mu^{+}\in P_{+}$, we have
\begin{align}
 \frac{\langle F_{\mu},F_{\mu}\rangle_{k}}
      {\langle F_{\mu^{+}},F_{\mu^{+}}\rangle_{k}}
 =\prod_{\alpha^{\vee}\in R_{w_{\mu}}^{\vee}}
  \frac{\langle\alpha^{\vee},\mu^{+}+\rho_{k}\rangle^{2}}
       {\langle\alpha^{\vee},\mu^{+}+\rho_{k}\rangle^{2}-k_{\alpha}^{2}}.
\end{align}
\end{prop}
\begin{proof}
For a reduced expression 
$w_{\mu}=s_{i_{l}}\cdots s_{i_{2}}s_{i_{1}}$,
we have
\begin{align}
 \hat{K}_{w_{\mu}}^{*}\hat{K}_{w_{\mu}}
 &=\hat{K}_{i_{l}}^{*}\cdots \hat{K}_{i_{2}}^{*}\hat{K}_{i_{1}}^{*}
   \hat{K}_{i_{1}}\hat{K}_{i_{2}}\cdots \hat{K}_{i_{l}} \nn\\
 &=\hat{K}_{i_{l}}^{*}\cdots \hat{K}_{i_{2}}^{*}
   \hat{N}(\alpha_{i_{1}}^{\vee})
   \hat{K}_{i_{2}}\cdots \hat{K}_{i_{l}} \nn\\
 &=\prod_{\alpha^{\vee}\in R_{w_{\mu}}^{\vee}}
   \!\! \hat{N}(\alpha^{\vee}). \nn
\end{align}
{}From the Rodrigues formula \myref{eq:nonsym-HO-Rodrigues2},
we calculate the scalar product as follows:
\begin{align}
 \langle F_{\mu},F_{\mu}\rangle_{k}
 &=\langle c_{w_{\mu}}^{-1}\hat{K}_{w_{\mu}}F_{\mu^{+}},
   c_{w_{\mu}}^{-1}\hat{K}_{w_{\mu}}F_{\mu^{+}}\rangle_{k} \nn\\
 &=(c_{w_{\mu}})^{-2}
   \langle F_{\mu^{+}},\hat{K}_{w_{\mu}}^{*}
   \hat{K}_{w_{\mu}}F_{\mu^{+}}\rangle_{k} \nn\\
 &=(c_{w_{\mu}})^{-2}
   \!\!\prod_{\alpha^{\vee}\in R_{w_{\mu}}^{\vee}}\!\!
   \langle F_{\mu^{+}},\hat{N}(\alpha^{\vee})F_{\mu^{+}}\rangle_{k} \nn\\
 &=(c_{w_{\mu}})^{-2}
   \langle F_{\mu^{+}},F_{\mu^{+}}\rangle_{k}
   \!\!\prod_{\alpha^{\vee}\in R_{w_{\mu}}^{\vee}}\!\!
   (\langle\alpha^{\vee},\mu^{+}+\rho_{k}\rangle^{2}-k_{\alpha}^{2}).\nn
\end{align}
\end{proof}

\section{Bosonic and Fermionic Eigenstates for Generalized
Sutherland models}

The generalized Sutherland models associated with arbitrary reduced
root systems are given by
\begin{align}
 \label{eq:g-Sutherland_1}
  H_{\rm S} 
  &\Define
   \sum_{i\in I}\partial^{\Lambda_{i}^{\vee}}\partial^{\alpha_{i}}
  -\sum_{\alpha\in R_{+}}
   \frac{\langle\alpha,\alpha\rangle}
        {(x^{\alpha/2}-x^{-\alpha/2})^{2}}
        k_{\alpha}(k_{\alpha}-s_{\alpha}). 
\end{align}
Using the variables $\{t\in T\}$, this is rewritten as
\begin{align}
  H_{\rm S}(t)
  =-\triangle
   +\frac{1}{4}
    \sum_{\alpha\in R_{+}}\frac{\langle\alpha,\alpha\rangle}
   {\sin^{2}(\langle t,\alpha\rangle/2)}k_{\alpha}(k_{\alpha}-s_{\alpha}),
\end{align}
where $\triangle$ is the Laplacian on $T$.
The Hamiltonian has so-called exchange terms
\cite{Baker-Forrester_97NPB}.
If we consider the bosonic (or fermionic) eigenstates, i.e.,
we restrict the operand of $H_{\rm S}$
to the $W$-symmetric (or $W$-anti-asymmetric) function space,
we can replace the exchange terms by $s_{\alpha}=1$ (or $-1$),
\begin{align}
 \label{eq:g-Sutherland_2}
  H_{\rm S}^{(B,F)}
  &=\sum_{i\in I}\partial^{\Lambda_{i}^{\vee}}\partial^{\alpha_{i}}
  -\sum_{\alpha\in R_{+}}
   \frac{\langle\alpha,\alpha\rangle}
        {(x^{\alpha/2}-x^{-\alpha/2})^{2}}
        k_{\alpha}(k_{\alpha}\mp 1). 
\end{align}
Through the similarity transformation by a $W$-symmetric function
\begin{align}
  \phi_{k}\Define
  \prod_{\alpha\in R}|1-x^{\alpha}|^{k_{\alpha}/2},
\end{align}
we obtain
\begin{align}
 \phi_{k}^{-1}\circ H_{\rm S}\circ\phi_{k}
 =\sum_{i\in I}
  \hat{D}^{\Lambda_{i}^{\vee}}\hat{D}^{\alpha_{i}}.
\end{align}
Hence we see that $H_{\rm S}$ has the eigenvectors
in $\mathbb{C}[P]\phi_{k}$ written by products of
the nonsymmetric Heckman-Opdam polynomial $F_{\mu}$ and $\phi_{k}$.
We note that $\phi_{k}$ corresponds to the ground state wave function
for the bosonic Hamiltonian $H_{\rm S}^{(B)}$.
Since the nonsymmetric eigenstates expressed by products of
$\phi_{k}$ and the nonsymmetric Heckman-Opdam polynomials
with weights lying in the same $W$-orbit
have the same eigenvalues of $H_{\rm S}$,
\begin{align}
 H_{\rm S}(\phi_{k}F_{\mu})
 &=\sum_{i\in I}\langle\Lambda_{i}^{\vee},\mu+\rho_{k}(\mu)\rangle
               \langle\alpha_{i},\mu+\rho_{k}(\mu)\rangle(\phi_{k}F_{\mu})
   \nn\\
 &=\langle\mu+\rho_{k}(\mu),\mu+\rho_{k}(\mu)\rangle(\phi_{k}F_{\mu})
   \nn\\
 &=\langle\mu^{+}+\rho_{k},\mu^{+}+\rho_{k}\rangle(\phi_{k}F_{\mu}),
\end{align}
we can take any linear combinations of $\phi_{k}F_{\mu}$
with weights lying in a $W$-orbit as eigenvectors of $H_{\rm S}$
(see \cite{Baker-Forrester_97NPB} for type $A$).
In what follows, we construct the bosonic and fermionic eigenstates
of the generalized Sutherland models
$H_{\rm S}^{(B,F)}$ from the $W$-symmetrized and $W$-anti-symmetrized
nonsymmetric Heckman-Opdam polynomials, respectively.

\begin{thm}
Let $J_{\mu}^{+}, (\mu\in P_{+})$ and 
$J_{\mu}^{-}, (\mu\in P_{++})$ be the following 
linear combinations of the nonsymmetric Heckman-Opdam polynomials
$F_{\tilde{\mu}}, (\tilde{\mu}\in W(\mu))$:
\begin{subequations}
\begin{align}
 &J_{\mu}^{\pm}=\sum_{\tilde{\mu}\in W(\mu)}
  b_{\mu\tilde{\mu}}^{\pm}F_{\tilde{\mu}}, 
\end{align}
where
\begin{align}
  b_{\mu\tilde{\mu}}^{\pm}
  =\prod_{\alpha^{\vee}\in R_{w_{\tilde{\mu}}}^{\vee}}
  \pm\frac{\langle\alpha^{\vee},\mu+\rho_{k}\rangle\mp k_{\alpha}}
       {\langle\alpha^{\vee},\mu+\rho_{k}\rangle}.
\end{align}
\end{subequations}
The polynomials $J_{\mu}^{+}, (\mu\in P_{+})$ and
$J_{\mu}^{-}, (\mu\in P_{++})$ are elements of 
$\mathbb{C}[P]^{\pm W}$
and called the symmetric and
the anti-symmetric Heckman-Opdam polynomials respectively.
\end{thm}
\begin{proof}
 It is sufficient to require the conditions 
 $s_{i}J_{\mu}^{\pm}=\pm J_{\mu}^{\pm}$ and
 $b_{\mu\mu}^{\pm}=1$ in order to determine 
 the coefficients $b_{\mu\tilde{\mu}}^{\pm}$
 such that $J_{\mu}^{\pm}\in\mathbb{C}[P]^{\pm W}$.
\end{proof}
The symmetric Heckman-Opdam polynomials of type $A$
are equivalent to the (symmetric) Jack polynomials.
Symmetrization of the nonsymmetric Jack polynomials 
was carried out by Baker and Forrester
\cite{Baker-Forrester_97q-alg}.
We remark that
their method with arm- and leg-lengths of the Young diagram
is essentially different from our approach.

We obtain the Rodrigues formulae for the symmetric
and the anti-symmetric Heckman-Opdam polynomials
$J_{\mu}^{\pm}$, 
($\mu\in P_{+}$ for $J_{\mu}^{+}$ and
 $\mu\in P_{++}$ for $J_{\mu}^{-}$),
\begin{align}
 \label{eq:sym-Rodrigues}
 J_{\mu}^{\pm}
 =\sum_{\tilde{\mu}\in W(\mu)}b_{\mu\tilde{\mu}}^{\pm}
   c_{w_{\tilde{\mu}}}^{-1} c_{\mu}^{-1}
   \hat{K}_{w_{\tilde{\mu}}}\hat{B}_{\mu}F_{0}.
\end{align}
As a result, we find the bosonic and the fermionic eigenstates 
$\phi_{\mu}^{(B,F)}$, ($\mu\in P_{+}$ for $\phi_{\mu}^{(B)}$ and
 $\mu\in P_{++}$ for $\phi_{\mu}^{(F)}$)
for the generalized Sutherland models 
$H_{\rm S}^{(B,F)}$,
\begin{align}
 &\phi_{\mu}^{(B,F)}
 =\phi_{k}
  \sum_{\tilde{\mu}\in W(\mu)}b_{\mu\tilde{\mu}}^{\pm}
  c_{w_{\tilde{\mu}}}^{-1} c_{\mu}^{-1}
  \hat{K}_{w_{\tilde{\mu}}}\hat{B}_{\mu}F_{0}, \nn\\
 &H_{\rm S}^{(B,F)}\phi_{\mu}^{(B, F)}
  =\langle\mu+\rho_{k},\mu+\rho_{k}\rangle\phi_{\mu}^{(B, F)},
\end{align}
respectively.
Lapointe and Vinet obtained 
the Rodrigues formulae for the Jack polynomials
\cite{Lapointe-Vinet_95IMRN}.
The relation between our formulae \myref{eq:sym-Rodrigues}
and theirs has not been clarified.

We proceed to the evaluation of
square norms of the eigenstates $\phi^{(B, F)}$,
\begin{align}
 \label{eq:BF-norm}
 \|\phi_{\mu}^{(B,F)}\|^{2}
 &=\int_{T}|\phi_{\mu}^{(B,F)}(t)|^{2} d \mu
  =\langle J_{\mu}^{\pm},J_{\mu}^{\pm}\rangle_{k}.
\end{align}
To prove a theorem, we need the following lemma:
\begin{lm}
 \label{lm:key-lemma2}
For $\mu\in P_{+}$, we have an identity,
\begin{align}
 \label{eq:key-identity}
 \sum_{\tilde{\mu}\in W(\mu)}
 \prod_{\alpha^{\vee}\in R_{w_{\tilde{\mu}}}^{\vee}}
   \frac{\langle\alpha^{\vee},\mu+\rho_{k}\rangle\mp k_{\alpha}}
        {\langle\alpha^{\vee},\mu+\rho_{k}\rangle\pm k_{\alpha}}
 =\prod_{\alpha\in R_{+}}
   \frac{\langle\alpha^{\vee}\!,\mu+\rho_{k}\rangle}
        {\langle\alpha^{\vee}\!,\mu+\rho_{k}\rangle\pm k_{\alpha}}.
\end{align}
\end{lm}
The identity is proved by using an expression of
the Poincar\'{e} polynomials 
\cite{Macdonald_72MathAnn,Nishino-Komori_00preprint}.
We shall show a proof in Appendix A.

\begin{thm}
Let $\mu\in P_{+}$ for $J_{\mu}^{+}$
and let $\mu\in P_{++}$ for $J_{\mu}^{-}$. We have
\begin{align}
\label{eq:HO_inner-prod-id}
 \langle J_{\mu}^{\pm},J_{\nu}^{\pm}\rangle_{k}
 =\delta_{\mu\nu}\prod_{\alpha\in R_{+}}
  \frac{\Gamma(\langle\alpha^{\vee},\mu+\rho_{k}\rangle\pm k_{\alpha})
        \Gamma(\langle\alpha^{\vee},\mu+\rho_{k}\rangle\mp k_{\alpha}+1)}
        {\Gamma(\langle\alpha^{\vee},\mu+\rho_{k}\rangle)
        \Gamma(\langle\alpha^{\vee},\mu+\rho_{k}\rangle+1)}.
\end{align}
\end{thm}
\begin{proof}
The orthogonality for $\mu\neq \nu$ is straightforward
from that of the nonsymmetric Heckman-Opdam polynomials
\myref{eq:nonsym-HO-orth}.
We have
\begin{align}
 \langle J_{\mu}^{\pm},J_{\mu}^{\pm}\rangle_{k}
 &=\sum_{\tilde{\mu}\in W(\mu)}
   (b_{\mu\tilde{\mu}}^{\pm})^{2}
       \langle F_{\tilde{\mu}},F_{\tilde{\mu}}\rangle_{k} \nn\\
 &=\sum_{\tilde{\mu}\in W(\mu)}
   (b_{\mu\tilde{\mu}}^{\pm})^{2}
  \frac{\langle F_{\tilde{\mu}},F_{\tilde{\mu}}\rangle_{k}}
       {\langle F_{\mu},F_{\mu}\rangle_{k}}
       \langle F_{\mu},F_{\mu}\rangle_{k} \nn\\
 &=\sum_{\tilde{\mu}\in W(\mu)}
   \prod_{\alpha^{\vee}\in R_{w_{\tilde{\mu}}}^{\vee}}
   \frac{\langle\alpha^{\vee},\mu+\rho_{k}\rangle\mp k_{\alpha}}
        {\langle\alpha^{\vee},\mu+\rho_{k}\rangle\pm k_{\alpha}}
        \langle F_{\mu},F_{\mu}\rangle_{k} \nn\\
 &=\prod_{\alpha\in R_{+}}
   \frac{\langle\alpha^{\vee}\!,\mu+\rho_{k}\rangle}
        {\langle\alpha^{\vee}\!,\mu+\rho_{k}\rangle\pm k_{\alpha}}
   \langle F_{\mu},F_{\mu}\rangle_{k}, \nn
\end{align}
where the last equality follows from Lemma \ref{lm:key-lemma2}.  
\end{proof}

\begin{cor}
For $k_{\alpha}\in \mathbb{N}$, 
we have
\begin{align}
 \label{eq:shift}
 \langle J_{\mu}^{\pm},J_{\nu}^{\pm}\rangle_{k}
 =\delta_{\mu,\nu}\prod_{\alpha\in R_{+}}\prod_{i=1}^{k_{\alpha}-1}
  \frac{\langle\alpha^{\vee},\mu+\rho_{k}\rangle\pm i}
       {\langle\alpha^{\vee},\mu+\rho_{k}\rangle\mp i}.
\end{align}
\end{cor}
We remark that
the inner products \myref{eq:shift} were first proved by use of
Opdam's shift operators \cite{Opdam_89InvMath}.
From \myref{eq:BF-norm} and \myref{eq:HO_inner-prod-id},
we obtain square norms of the eigenstates 
$\phi^{(B, F)}_{\mu}$
for the generalized Sutherland models,
\begin{align}
 \|\phi_{\mu}^{(B,F)}\|^{2}
 =\prod_{\alpha\in R_{+}}
  \frac{\Gamma(\langle\alpha^{\vee},\mu+\rho_{k}\rangle\pm k_{\alpha})
        \Gamma(\langle\alpha^{\vee},\mu+\rho_{k}\rangle\mp k_{\alpha}+1)}
        {\Gamma(\langle\alpha^{\vee},\mu+\rho_{k}\rangle)
        \Gamma(\langle\alpha^{\vee},\mu+\rho_{k}\rangle+1)}.
\end{align}

\section{Concluding Remarks}
We summarize the results in this paper.
First we have presented the Rodrigues formulae for
the nonsymmetric Heckman-Opdam polynomials which correspond
to the nonsymmetric basis of the generalized Sutherland models
with exchange terms.
Their square norms are evaluated in an algebraic manner.
Second through $W$-symmetrization and $W$-anti-symmetrization
of the nonsymmetric Heckman-Opdam polynomials
we have constructed the bosonic and fermionic eigenstates
of the generalized Sutherland models with arbitrary reduced root systems
respectively.
The square norms of the eigenstates
are calculated from their nonsymmetric counterparts
through an expression of the Poincar\'{e} polynomials.

We consider
some interesting applications and extensions of our method.
The generalized Sutherland models we have studied in this paper
do not include the models associated with the $BC_{N}$-type
nonreduced root system.
Since we have already obtained the Rodrigues formulae
for the nonsymmetric Heckman-Opdam polynomials of type $BC_{N}$
\cite{Nishino-Ujino_00NPB,Ujino-Nishino_99proc},
the extension should be straightforward.
And it is known that there exists the symmetric orthogonal basis
for the Calogero model which describes many particles 
with inverse-square interactions in a harmonic well 
\cite{Ujino-Wadati_97JPSJ,Ujino-Wadati_99JPSJ,Nishino-Ujino_99aJPSJ,
Ujino-Nishino_99proc}.
We have already confirmed that our method can be applied to 
the eigenstates of the Calogero model. 
The detail of the analysis will be reported elsewhere.

\begin{appendix}

\section{Proof of Lemma \ref{lm:key-lemma2}}
We present a proof of Lemma \ref{lm:key-lemma2}
following our previous paper \cite{Nishino-Komori_00preprint}.

The Poincar\'{e} polynomials associated with the Weyl group
\cite{Humphreys}
are given by
\begin{align}
 W(t)
 =\sum_{w\in W}
   \prod_{\alpha\in R_{w}}t_{\alpha},
\end{align}
where $\{t_{\alpha}|\alpha\in R\}$
are $W$-invariant indeterminates, i.e., $t_{\alpha}=t_{w(\alpha)}$
for $w\in W$.
We denote  by $\mathbb{K}$ 
the field of rational functions over $\mathbb{C}$
in square-roots of indeterminates $\{t_{\alpha}\}$.
To investigate the Poincar\'{e} polynomials,
Macdonald proved the following identity \cite{Macdonald_72MathAnn}:
\begin{thm}[I.~G.~Macdonald]
\begin{align}
 W(t)=
 \sum_{w\in W}\prod_{\alpha\in R_{+}}
  \frac{1-t_{\alpha}x^{w(\alpha^{\vee})}}{1-x^{w(\alpha^{\vee})}}.
 \label{eq:Poincare_1}
\end{align}
\end{thm}

\begin{lm}
\label{lm:key-lemma}
Let $\mu\in P_{+}$. We have
\begin{align}
 \label{eq:w-sum}
 \sum_{\tilde{\mu}\in W(\mu)}
 \prod_{\alpha^{\vee}\in R_{w_{\tilde{\mu}}}^{\vee}}
   \frac{t_{\alpha}(1-t_{\alpha}^{-1}
         q^{\pm \langle\alpha^{\vee},\mu+\rho_{k}\rangle})}
        {1-t_{\alpha}
         q^{\pm \langle\alpha^{\vee},\mu+\rho_{k}\rangle}}
 =W(t)\prod_{\alpha\in R_{+}}
   \frac{1-q^{\pm \langle\alpha^{\vee}\!,\mu+\rho_{k}\rangle}}
        {1-t_{\alpha}q^{\pm \langle\alpha^{\vee}\!,\mu+\rho_{k}\rangle}}.
\end{align}
\end{lm}
\begin{proof}
 There exists a $\mathbb{K}$-homomorphism 
$\varphi : 
 \mathbb{K}[Q^{\vee}]\rightarrow \mathbb{K}$ defined by
\[
 \varphi:
 x^{\alpha_{i}^{\vee}}
 \mapsto q^{\pm\langle\alpha_{i}^{\vee},\mu+\rho_{k}\rangle},\,
 \text{ for }\, i\in I.
\]
Since 
$W(t)\in \mathbb{K}[Q^{\vee}]$ does not
depend on $\{x^{\alpha_{i}^{\vee}}\}$ as \myref{eq:Poincare_1}, we have
\begin{align} 
 &\varphi\big(W(t)\big)=W(t) \nn\\
 &=\sum_{w\in W}\prod_{\alpha\in R_{+}}
   \varphi\Big(
   \frac{1-t_{\alpha}x^{w(\alpha^{\vee})}}{1-x^{w(\alpha^{\vee})}}
   \Big) \nn\\
 &=\sum_{w\in W}\prod_{\alpha\in R_{+}}
   \frac{1-t_{\alpha}q^{\pm\langle w(\alpha^{\vee}),\mu+\rho_{k}\rangle}}
        {1-q^{\pm\langle w(\alpha^{\vee}),\mu+\rho_{k}\rangle}} \nn\\
 &=\frac{\displaystyle{
       \sum_{w\in W}
       \prod_{\alpha^{\vee}\in R_{w}^{\vee}}
       (t_{\alpha}\!-\!
        q^{\pm\langle\alpha^{\vee},\mu+\rho_{k}\rangle})\!\!\!\!
       \prod_{\alpha^{\vee}\in R_{+}^{\vee}\setminus R_{w}^{\vee}}\!\!\!\!
       (1\!-\! t_{\alpha}
        q^{\pm\langle\alpha^{\vee},\mu+\rho_{k}\rangle})
      }}
      {\displaystyle{
       \prod_{\alpha\in R_{+}}
       (1\!-\! q^{\pm\langle\alpha^{\vee},\mu+\rho_{k}\rangle})
      }}. \nn
\end{align}
Thus we obtain the following relation:
\begin{align}
 \label{eq:relation}
 \sum_{w\in W}\prod_{\alpha^{\vee}\in R_{w}^{\vee}}
   \frac{t_{\alpha}(1-t_{\alpha}^{-1}
         q^{\pm\langle\alpha^{\vee},\mu+\rho_{k}\rangle})}
        {1-t_{\alpha}
         q^{\pm\langle\alpha^{\vee},\mu+\rho_{k}\rangle}}
 =W(t)\prod_{\alpha\in R_{+}}
   \frac{1-q^{\pm\langle\alpha^{\vee}\!,\mu+\rho_{k}\rangle}}
        {1-t_{\alpha}q^{\pm\langle\alpha^{\vee}\!,\mu+\rho_{k}\rangle}}.
\end{align}
We show that the sum on the left-hand side of the above equation
can be replaced by the sum on $\tilde{\mu}\in W(\mu)$.
Consider the isotropy group $W_{\mu}=\{w\in W|w(\mu)=\mu\}$
for the dominant weight $\mu\in P_{+}$ 
($W_{\mu}=\{1\}$ for $\mu\in P_{++}$).
Since an element $w\in W_{\mu}\setminus\{1\}$ 
can be written by a product of
simple reflections fixing $\mu$, 
$\{s_{i}|i\in J\subset I\}$ (see \cite{Humphreys}),
there exists at least one simple root $\alpha_{i}^{\vee}\in \Pi^{\vee}$
associated with the reflection $s_{i}$ in the set $R_{w}^{\vee}$.
Hence, for $w\in W_{\mu}\setminus\{1\}$, we have
\begin{align}
 \prod_{\alpha^{\vee}\in R_{w}^{\vee}}
 t_{\alpha}(1-t_{\alpha}^{-1}
               q^{\langle\alpha^{\vee},\mu+\rho_{k}\rangle})
 &=t_{i}(1-t_{i}^{-1}
               q^{\langle\alpha_{i}^{\vee},\rho_{k}\rangle})
  \prod_{\alpha^{\vee}\in R_{w}^{\vee}\setminus \{\alpha_{i}^{\vee}\}}
  t_{\alpha}(1-t_{\alpha}^{-1}
               q^{\langle\alpha^{\vee},\mu+\rho_{k}\rangle}) \nn\\
 &=t_{i}(1-t_{i}^{-1}t_{i})
  \prod_{\alpha^{\vee}\in R_{w}^{\vee}\setminus \{\alpha_{i}^{\vee}\}}
  t_{\alpha}(1-t_{\alpha}^{-1}
               q^{\langle\alpha^{\vee},\mu+\rho_{k}\rangle})=0. \nn
\end{align}
Define $W^{\mu}\Define \{w\in W|l(ws_{i})>l(w)\, 
\text{ for all }\, i\in J\}$.
For $w\in W$, there is a unique $u\in W^{\mu}$ 
and a unique $v\in W_{\mu}$ such that $w=uv$.
We obtain the above lemma since the sum on $w\in W$ on
the left-hand side of \myref{eq:relation}
can be replaced by that on $w\in W^{\mu}$
which is equivalent to that on $\Tilde{\mu}\in W(\mu)$.
\end{proof}

In the formal limit $q\rightarrow 1$ under the restriction
$t_{\alpha}=q^{k_{\alpha}}$, we have the relation 
\myref{eq:key-identity} in Lemma \ref{lm:key-lemma2}.

\end{appendix}

\section*{Acknowledgments}
The authors would like to thank Dr.~H.~Ujino and Dr.~Y.~Komori
for valuable discussions. One of the authors (AN) would like to express
his gratitude to Prof.~M.~J.~Ablowitz, Prof.~H.~Segur, Prof.~L.~Vinet and
their colleagues in Department of Applied Mathematics, University of
Colorado and Centre de Recherches Math\'{e}matique, Universit\'{e} de
Montr\'{e}al for giving kind hospitality and helpful comments during
his stay in Boulder and Montreal.


\end{document}